\documentclass[12pt]{article}
\usepackage{epsf}
\usepackage{amsmath}

\usepackage{graphics}
\usepackage{cite}

\begin{titlepage}

\begin{document}

\hfill December 2020\\

\hfill Revised\\

\begin{center}

{\bf \Large Swampland Conjectures and Cosmological Expansion \\
}
\vspace{0.5cm}
{\bf Claudio Corian\`{o}
\footnote{claudio.coriano@le.infn.it} 
and Paul H. Frampton
\footnote{ paul.h.frampton@gmail.com} 
\\  }
\vspace{0.5cm}
{\it Dipartimento di Matematica e Fisica "Ennio De Giorgi",\\ 
Universit\`{a} del Salento and INFN-Lecce,\\ Via Arnesano, 73100 Lecce, Italy
}

\vspace{1.0in}

\bigskip

\begin{abstract}
\noindent
Swampland conjectures (SCs) of string theory
require that a constant cosmological
constant $\Lambda$ be replaced by a time-dependent
scalar-field quintessence with constrained parameters.
The constraints limit the duration of the present expansion era because, 
although the SCs may be fulfilled
at the present time, they will be violated at a finite time in the future
allowing only an order-one number of e-foldings. 
In contrast, cyclic cosmology requires $\sim94$ e-foldings of the
present universe before turnaround from expansion to contraction.
This presents a dilemma to the original SCs. 
One possibility is that one of the SCs, the range conjecture, be significantly weakened.
A second possibility, difficult to believe, is that cyclic cosmology vastly overestimates
the number of e-foldings.
A third possibility, which is the least disfavoured, is that string theory is 
not the correct theory of quantum gravity.

\end{abstract}

\end{center}
\end{titlepage}

\section{Introduction} 

\noindent
An interesting question is whether string theory is the correct theory
of quantum gravity. Popular books often state that the motivation
for string theory is to reconcile general relativity with quantum mechanics,
especially at microscopic distances. This is not the whole story. String 
theory was invented originally to describe strong nuclear interactions
and was, and remains, quite successful in that goal. String theory is
an impressive mathematical framework which has inspired significant
progress in pure mathematics.

\bigskip

\noindent
String theory is over fifty years old and
it might be expected that it could be even more 
decades before there exist relevant experimental
or observational data on quantum gravity that could enable
an informed response to our
initial question: is string theory the correct theory of quantum gravity?

\bigskip

\noindent
But if the answer to this question is negative, we may be able to decide
much sooner. Extraordinarily strong swampland conjectures
\cite{Vafa2005,OoguriVafa2006,OOSV,OPSV} have been
proposed, so strong that preclusion of agreement with existing data is
feasible. We could consider either experiments testing particle
theory at colliders or astronomical observations which confront theoretical cosmology.
In the present article, we favour theoretical cosmology as the better testing ground,
particularly the future of the universe. This may seem paradoxical
because no observations are possible, but that is our contention. 

\bigskip

\noindent
It is of broad interest to understand whether the present expansion of the universe
will last for an infinite time as in the $ \Lambda CDM $ model with constant $\Lambda$
or whether the present expansion will end at a future finite time, to be followed by a contraction
era as in an infinitely cyclic cosmology which can provide a more satisfactory
explanation of how time never began at a finite past time. Although
the future behaviour is not directly observable, its mathematical description can lead
to testable predictions at the present time concerning {\it e.g.} the equation of state of the dark
energy.

\bigskip

\noindent
This question will be studied in the present article within the most developed cyclic cosmology
invented\cite{BF} in 2007 in which the surprisingly accurate estimate according to a 
calculation by one of us\cite{PHF} is that the end of
the present expansion era will occur at a turnaround time, $t_T=1.3$Ty, when expansion ends and contraction begins, derived by the matching of expansion and contraction scale factors
necessary for a consistent cyclic cosmology to be of infinite duration in both past and future.

\bigskip

\noindent
Although the swampland conjectures (SCs) concerning string theory were
first enunciated in 2005 and 2006 by  Ooguri and Vafa\cite{Vafa2005,OoguriVafa2006}, subsequently
modified in \cite{OOSV,OPSV} the implications
of these SCs for cosmology were first carefully considered only in 2018 {\it e.g.} \cite{AOSV}. Implications of the SCs for particle theory have also been
discussed \cite{DHW,CCS,C,March-Russell} but, while that is another very interesting direction, in the
present article we shall focus on predictions of the SCs for cyclic cosmology.

\bigskip

\noindent
The SCs are tied to the assumption that string theory, in which we include M-theory
and F-theory, is the correct theory of quantum gravity and were suggested based on
that optimistic assumption. In the present article, we adopt the SCs as being correct
and study what they can tell us about the future of the
universe.

\bigskip

\noindent
It is worth mentioning that the minimal standard model of particle theory
with only one scalar doublet violates the SCs and that the standard $ \Lambda CDM$
cosmological model with $\Lambda$ constant violates the SCs. Thus the SCs are
very powerful and do not respect long-held prejudices. This is what makes the
SCs so remarkably interesting that they merit further study. An additional scalar field $\phi$
must be added to accommodate the two very successful theories of particle theory
and cosmology.

\bigskip

\noindent
The swampland conjectures which have been deemed necessary, to ensure
that a low-energy effective field theory have an ultra-violet completion within
string theory and so belong to the string landscape rather than the very much bigger
swampland, are that the scalar field $\phi$ and its potential $V(\phi)$ satisfy
two swampland conjectures, SC1 and SC2 as follows:

\bigskip

\noindent
{\it SC1:~~~~~ The range traversed by $\phi$ in field space is bounded by
$\Delta \sim O(1)$ in reduced Planck units}.

\bigskip

\noindent
{\it SC2: ~~~~~$ \left( \frac{| \nabla_{\phi} V|}{V} \right)  \geq c$ or  $\min (\nabla_i\nabla_j V) \leq  -c'$ with $c, c'>0$, both $O(1)$ in reduced Planck units,
and where $\min (\nabla_i\nabla_j V)$ is the minimum eigenvalue of the Hessian $\nabla_i\nabla_jV$ in an orthonormal frame.}.

\bigskip

\noindent
These conjectures are not rigorously proven in string theory but are supported by all string theory
examples so far studied assiduously and hence a working hypothesis is
to assume that they are necessary as well as sufficient for a successful UV completion.
The conjectures {\it ut supra} are sometimes
called the ``range" constraint and the ``de Sitter" constraint, respectively, but we shall denote them
{\it ut infra} simply by SC1 and SC2.

\bigskip

\noindent
Assuming SC1 and SC2 of string theory is not good news for the principal theories of
inflation \cite{Storobinsky,Guth,Linde,Albrecht} as shown in \cite{AOSV}
whose arguments we shall accept and not attempt to improve upon. This may 
disappoint the inflation community but does lend support for cyclic cosmology
in which a cosmic bounce from contraction to expansion could provide an
explanation, alternative to inflation, for the flatness and other issues. 
We now discuss dark energy represented by a quintessence scalar field $\phi$.

\section{Quintessence}

\noindent
We start from the action
\begin{equation}
S = \int d^4x \sqrt{-g} \left[ \frac{1}{2} M_{Pl}^2 R - \frac{1}{2} g^{\mu\nu}\partial_{\mu}\phi\partial_{\nu}
\phi - V(\phi) \right] + S_m,
\label{action}
\end{equation}
in which $M_{Pl}$ is the Planck mass, $\phi$ is a scalar field and $S_m$ is the matter action.

\bigskip

\noindent
We assume a FLRW metric with scale factor $a(t)$ normalised to $a(t_0)=1$ at the present time
and Hubble parameter $H(t) = \dot{a}(t)/a(t)$
which has the present value $H(t_0)=H_0$. We define the density, pressure of dark
energy as $\rho_{\phi}$, $p_{\phi}$ so that the dark energy equation of state $\omega$ is
\begin{equation}
\omega = \left( \frac{p_{\phi}}{\rho_{\phi}} \right) = \left( \frac{ \frac{1}{2}\dot{\phi}^2 - V(\phi)}{ \frac{1}{2}\dot{\phi}^2 + V(\phi)} \right),
\label{omega}
\end{equation}
since $p_{\phi} = \frac{1}{2}\dot{\phi}^2 - V(\phi)$ and $\rho_{\phi}=\frac{1}{2}\dot{\phi}^2 + V(\phi)$.

\bigskip

\noindent
The continuity equation is
\begin{equation}
\dot{\rho}_{\phi} + 3 H (\rho_{\phi}+p_{\phi}) = 0.
\label{continuity}
\end{equation}
Differentiating this and defining $V_{,\phi} = dV/d\phi$ we find that
\begin{equation}
\ddot{\phi} + 3 H \dot{\phi} + V_{,\phi} = 0.
\label{continuity2}
\end{equation}

\bigskip

\noindent
The two equations of motion arising from the action in Eq.(\ref{action}) are
\begin{equation}
3 M_{Pl}^2 H^2 = \left( \frac{\dot{\phi}^2}{2} \right) + V(\phi) + \rho_m,
\label{EOM1}
\end{equation}

\bigskip

\noindent
and

\bigskip

\begin{equation}
2 M_{Pl}^2 \dot{H} = - \left[ \dot{\phi}^2 + (1 + \omega_m) \rho_m \right],
\label{EOM2}
\end{equation}
where $\omega_m$ is the equation of state $\omega_m = p_m/\rho_m$
corresponding to the matter described by $S_m$ in Eq.(\ref{action}).

\bigskip

\noindent
To calculate the time evolution of the scalar field, we shall find it useful to employ variables $x, y$ first introduced by
Copeland, Liddle and Wands\cite{Copeland}:

\begin{equation}
x = \left( \frac{\dot{\phi}}{\sqrt{6} M_{Pl} H} \right); ~~~~~ 
y = \left( \frac{\sqrt{V(\phi)}}{\sqrt{3} M_{Pl} H} \right).
\label{xy}
\end{equation}

\bigskip

\noindent
The fraction of the critical density contributed by the dark energy is
\begin{equation}
\Omega_{\phi} = \left( \frac{\rho_{\phi}}{3 M_{Pl}^2 H^2} \right) 
=  \left( \frac{\frac{1}{2}\dot{\phi}^2 + V(\phi)}{3 M_{Pl}^2 H^2} \right) 
= x^2 + y^2.
\label{OmegaPhi}
\end{equation}

\bigskip

\noindent
The matter density is
\begin{equation}
\Omega_m = \left( \frac{\rho_m}{3 M_{Pl}^2 H^2} \right) = 1 - \Omega_{\phi} = 1-x^2-y^2.
\label{OmegaM}
\end{equation}

\bigskip

\noindent
Noting that
\begin{equation}
x^2 - y^2 = \left( \frac{\frac{1}{2} \dot{\phi}^2 - V(\phi)}{3 M_{Pl}^2 H^2} \right) =
\left( \frac{p_{\phi}}{3 M_{Pl}^2 H^2} \right),
\label{x2minusy2}
\end{equation}
we find for the equation of state $\omega$ of the dark energy
\begin{equation}
\omega = \frac{p_{\phi}}{\rho_{\phi}} = \left( \frac{x^2-y^2}{x^2+y^2} \right).
\label{omega}
\end{equation}
Eqs.(\ref{OmegaPhi}) and (\ref{omega}) show how $x,y$ relate to
physical quantities.

\bigskip

\noindent
The next step is to calculate how $x,y$ evolve with cosmic time for which
we use, as a convenient variable, the logarithm of the FLRW scale factor $N = \ln a$. Considering
first $x$, we have from Eq.(\ref{xy})
\begin{equation}
\frac{dx}{dN} = \frac{1}{\sqrt{6} M_{Pl}} \frac{d}{dN} \left( \frac{\dot{\phi}}{H} \right)
=  \frac{1}{\sqrt{6} M_{Pl}}\left(\frac{1}{H} \frac{d\dot{\phi}}{dN} - \frac{\dot{\phi}}{H^2}\frac{dH}{dN}
\right).
\label{dxdN}
\end{equation}
In the first term of Eq.(\ref{dxdN}) we use $d\dot{\phi}/dN = H^{-1}\ddot{\phi}$, and define
\begin{equation}
\lambda \equiv - M_{Pl} \frac{V_{,\phi}}{V}'
\label{lambda}
\end{equation}
then use Eq.(\ref{continuity2}) to rewrite this term as $-3x + \lambda y^2 \sqrt{6}/2$.

\newpage

\noindent
In the second term of Eq.(\ref{dxdN}), we use $dH/dN = \dot{H}/H$ and 
$\dot{H}/H^2 = 3x^2 +\frac{3}{2}(1+\omega_m)(1-x^2-y^2)$, to arrive at 
$\frac{3}{2} x \left[ (1-\omega_m)x^2 + (1 + \omega_m)(1-y^2) \right]$. 

\bigskip

\noindent
The time
dependence of $x$ is therefore
\begin{equation}
\frac{dx}{dN} = -3x + \frac{\sqrt{6}}{2} \lambda y^2 + \frac{3}{2} \left[
(1-\omega_m) x^2 + (1+\omega_m)(1-y^2) \right].
\label{finaldxdN}
\end{equation}

\bigskip

\noindent
Let us consider the time evolution of $y$ defined in Eq.(\ref{xy}) 
\begin{equation}
\frac{dy}{dN} = \frac{1}{\sqrt{3}M_{Pl}} \frac{d}{dN} \left( \frac{\sqrt{V(\phi)}}{H} \right)
= \frac{1}{3M_{Pl} H^2} \left( H \frac{d}{dN} \sqrt{V(\phi)} - \sqrt{V(\phi)} \frac{dH}{dN} \right).
\label{dydN}
\end{equation}

\bigskip

\noindent
The first term in Eq.(\ref{dydN}) is readily shown to equal $-\frac{\sqrt{6}}{2}\lambda xy$,
using Eqs. (\ref{xy}) and(\ref{lambda}). In the second term of Eq.(\ref{dydN}),
we use $dH/dN=\dot{H}/H$ and $\dot{H}/H^2= -3x^2-\frac{3}{2}(1+\omega_m)
(1-x^2-y^2)$, to arrive at
\begin{equation}
\frac{dy}{dN} = \frac{\sqrt{6}}{2} \lambda xy + \frac{3}{2}y 
\left[(1-\omega_m)x^2 + (1+\omega_m)(1-y^2) \right]
\label{finaldydN}
\end{equation}

\bigskip

\noindent
If we consider high redshift, observations of structure formation require
that $\Omega_{\phi} =x^2+y^2 \rightarrow 0$
and therefore in the xy-plane the time evolution locus begins in the past at the origin $(x, y) = (0, 0)$.
It evolves in time according to the coupled equations (\ref{finaldxdN}) and (\ref{finaldydN}).
This requires numerical analysis which was carried out in \cite{AOSV}, see therein especially
Figs. 1(a) and 1(b).

\bigskip

\noindent
In Fig 1(b) the locus (the blue line) moves from the origin to positive $(x, y)$ at the present
location where $\Omega_{\phi} (t=t_0) = \Omega_{\phi}^0 = x^2+y^2=0.7$.  Fig.1(a) expresses
the observational constraints on the dark energy equation of state $\omega(z)$
for the redshift range $0 < z < 1$.  In Fig. 1(b), if we veer to one side (right) of the
blue locus, we violate the equation of state $\omega$ constraint in Fig. 1(a), while if we go
to the other side (left) we find unphysical $\Omega_{\phi}$ because $x$ becomes negative
for high red shifts. Thus the blue line,
which is drawn for the maximum allowed value $c=0.6$ in SC2, is robust. This gives 
confidence in its extrapolation to future times. 

\bigskip

\noindent
What concerns us here is the extrapolation of the $(x, y)$ locus into the
future. This is the dotted blue line in Fig 1(b) and the crucial point is that,
although at the present time $t=t_0 = 13.8$Gy the swampland conjectures
are valid, SC1 will be violated at a finite time $t_T$ in the future.
We define $E_T$ as the number of e-foldings before $t=t_T$ {\it i.e.}
$(t_T - t_0) = E_T H^{-1}$. $E_T$ is estimated as
\begin{equation}
E_T \simeq \left( \frac{3 \Delta}{2 c \Omega_{\phi}^0} \right)
\label{Es}
\end{equation}
where $\Delta$ and $c$ are $O(1)$ constants appearing in
SC1 and SC2 respectively and $\Omega_{\phi} = 0.7$. A slightly
different estimate may be obtained from the modified Transplanckian Censorship
Conjecture recently proposed in 2019\cite{Vafa2019}, but the difference
is insufficient to change any conclusions of the present article.

\bigskip

\noindent
In summary, from Eq. (\ref{Es}), the swampland conjectures predict that
\begin{equation}
E_T = O(1),
\label{ETstring}
\end{equation}
implying that only a few e-foldings are permitted in the future cosmological
expansion before the turnaround to contraction.

\section{Cyclic Cosmology}

\noindent
To make this article self-contained, we include the calculation
of $t_T \simeq 1.3$Ty and, what is the same thing, $E_T \simeq 94$. Let us also
briefly review CBE cyclic cosmology introduced in \cite{BF} and pursued
in subsequent papers, because
assumptions made in the original paper
have since been weakened.

\bigskip

\noindent
CBE (= Comes Back Empty) cosmology is motivated by resolution
of the 1931 Tolman no-go theorem\cite{Tolman}
which pointed out that the second law of thermodynamics
and monotonic increase of entropy appeared at first sight,
and at that time even at second sight, 
contradictory to
infinitely cyclic cosmology because each period would be bigger
and longer then its predecessor. To avoid this no-go theorem, 
entropy must be periodically jettisoned, as is possible
only due to the existence of the dark energy unknown to Tolman.

\bigskip

\noindent
In the CBE model, entropy is jettisoned only at the turnaround from
expansion to contraction. At turnaround, after a very long period of
superluminal expansion, the universe fragments into a very large
number, measured in googols, of causal patches. Almost all
of these patches are empty meaning they contain no matter including black
holes. The remaining tiny fraction of causal patches, almost none, do contain
matter.

\bigskip

\noindent
Our contracting universe is not one of the causal patches containing
matter. Under contraction, black holes would
merge and grow. Matter would clump and create structure. Contraction
through phase transitions in reverse would violate the second law
of thermodynamics.
For all these reasons there will inevitably be a premature bounce
leading to a failed universe.

\bigskip

\noindent
 By contrast, a patch among the vast majority of patches which is
 empty can contract successfully. It contracts adiabatically with a
 time-reverse of the radiation era of expansion and with close to
 zero entropy, thus explaining why the present expansion began
 with extremely low entropy. The new much smaller scale factor $\hat{a}(t_T)$ for the 
contracting universe at turnaround is related to the scale factor
of the previous expansion $a(t_T)$ by 

\begin{equation}
\hat{a}(t_T)  = f a(t_T),
\label{f}
\end{equation}
where the coefficient $f << 1$ plays a role in the calculation of $t_T$,
{\it ut infra}. 

\bigskip

\noindent
Before doing that calculation, let us mention two assumptions
made in the original CBE paper \cite{BF} which subsequent work
has shown were not necessary.

\bigskip
 
\noindent
In \cite{BF}, it was assumed that the dark energy equation of state satisfied
$\omega<-1$, so-called phantom dark energy, because the inspiration came
from the Big Rip scenario\cite{Caldwell} in which time ends at a finite future
time.  However, this assumption is unnecessary and the CBE idea
about cyclic entropy works equally well for $\omega \geq -1$ so long
as there is lengthy superluminal expansion which will lead to the creation of
a {\it very large number} of causal patches before the turnaround.

\bigskip

\noindent
Also in \cite{BF}, it was assumed that there is inflation near to
the beginning of the expansion era. Such inflation is, however, not necessary
because the successful predictions of inflation 
can be reproduced by cyclic contraction to a bounce. For example, flatness is a natural
final state in a contracting FLRW metric and other issues like the horizon
problem and the scalar index may be accommodated. 

\bigskip

\noindent
It is worth pointing out a subtlety concerning the 
infinite past of an infinitely cyclic cosmology. There is an ambiguity in the
$t\rightarrow - \infty$ limit in that the number of universes either remains infinite 
or, perhaps surprisingly, can be finite {\it e.g.} one. To derive this 
result\cite{PHF2} requires the use
of set theory and transfinite numbers and can be of more interest
to mathematicians than to physicists.

\bigskip

\noindent
Let us begin by studying the present expansion era where
important cosmic times are when radiation domination is replaced by
matter domination ($t_m$), when matter domination is replaced by dark
energy domination ($t_{DE}$), the present age ($t_0$) and the future
turnaround time ($t_T$). For these we use the values \cite{PHF3}
\begin{eqnarray}
t_m & = &  47 ky,    \nonumber  \\
t_{DE} & = & 9.8 Gy,  \nonumber \\
t_0 & = & 13.8 Gy,   \nonumber \\
t_T & = & {\rm to ~ be ~ determined,} ~ ut ~  infra. \nonumber 
\label{times}
\end{eqnarray}
We must distinguish the radii of the introverse ($R_{IV}$)
and extroverse ($R_{EV}$) which, while they coincide at $t=t_{DE}$,
\begin{equation}
R_{IV}(t_{DE}) = R_{EV}(t_{DE}) = 39 Gly,
\label{equalityIVEV}
\end{equation}
for all later times satisfy $R_{EV}(t) > R_{IV}(t)$, for example at $t=t_0$
\begin{equation}
R_{EV}(t_0) = 52Gly; ~~~~~ R_{IV}(t_0) = 44 Gly. 
\label{t0RIVREV}
\end{equation}
\bigskip
\noindent
Taking cubes in Eq.(\ref{t0RIVREV}), the ratio of $EV$ to $IV$
volumes at present is
\begin{equation}
\left( \frac{V_{EV}(t_0)}{V_{IV}(t_0)} \right) = \left( \frac{ R_{EV}(t_0)^3}{R_{IV}(t_0)^3} \right) = 1.65.
\label{Volumes}
\end{equation}

\noindent
Eq.(\ref{Volumes}) implies that approximately  $40\%$ of the galaxies which were inside
the visible universe at $t=t_m=9.8$Gy have exited and the present visible universe
is surrounded by an extroverse containing hundreds of billions of galaxies rendered forever
invisible. This is an early precursor of the causal patch separation which will
take place at the turnaround time, $t=t_T$.

\bigskip

\noindent
Introverse means the same as visible universe, or particle horizon, and 
its radius $R_{IV}(t)$ given by
\begin{equation}
R_{IV}(t) = c \int_0^t \frac{dt}{a(t)},
\label{R(t)}
\end{equation}
where $a(t)$ characterises the expansion history of the universe and is the
scale factor in a flat FLRW metric
\begin{equation}
ds^2 = dt^2 - a(t)^2 \left[ dr^2 + r^2 (d\theta^2 + \sin^2 \theta d\phi^2) \right].
\label{metric}
\end{equation}

\bigskip

\noindent
As we have seen, the present value is
 $R_{IV}(t_0) = 44$Gly but, because of the finite speed of
light, it increases relatively slowly to its asymptotic value 
which is nearly reached already when
$t \sim 50$Gy
\begin{equation}
R_{IV}(t > 50 Gy) \simeq 58Gly.
\label{58Gly}
\end{equation}

\bigskip

\noindent
The extroverse radius $R_{EV}(t)$ expands exponentially until the turnaround
at $t=t_T$ when the number of causal patches can be estimated as 
\begin{equation}
N_T= \left( \frac{R_{EV}(t_T)^3}{R_{IV}(t_T)^3} \right) = \frac{1}{f^3},
\label{NTCBE}
\end{equation}
where $f$ was defined in Eq.(\ref{f}).
We require $N_T$ to be a very large number which means many googols 
because there are $\sim 10^{80}$ particles in the present 
extroverse and we need an overwhelming majority of empty causal
patches. This requirement of very large $N_T$ will be verified {\it a posteriori}.

\bigskip

\noindent
To calculate $a(t_T)$, we need to find $t_T$ from matching of the contraction
and expansion scale
factors, and use the value of $a(t_m=47ky)$ from \cite{PHF},
\begin{equation}
\hat{a}(t_m) = a(t_m) = 2.1 \times 10^{-4},
\label{matching}
\end{equation}
at the time $t=t_m$ because the radiation-dominated behaviour for
the expansion when $t<t_m$ matches the same behaviour of the entire
contraction. The matching in Eq.(\ref{matching}) is necessary
for a consistent infinite cyclicity. Provided $R_{IV}(t)$ is asymptotic,
to be justified {\it a posteriori}, we know that 
\begin{equation}
R_{EV}(t_T) = 58 Gly,
\label{REVT}
\end{equation} 
and therefore, since we use the normalisation $a(t_0)=1$, 
\begin{equation}
\hat{a}(t_T) = f a(t_T)  = \left( \frac{R_{IV}(t_T)}{R_{EV}(t_T)} \right) a(t_T)
= \left( \frac{R_{IV}(t_T)}{R_{EV}(t_0)} \right) = \left( 
\frac {58Gly}{52Gly} \right) = 1.11,
\end{equation}
independent of $t_T$, provided that $t_T > 50$Gy.
Now we use the time dependence appropriate to contraction of
the empty introverse
\begin{equation}
\hat{a}(t) = \hat{a}(t_T) \left( \frac{t}{t_T} \right)^{\frac{1}{2}},
\end{equation}
and the matching condition, Eq.(\ref{matching}), to calculate the turnaround time
\begin{equation}
t_T = \left(\frac{\hat{a}(t_T)}{\hat{a}(t_m)} \right) t_m 
= \left( \frac{1.11}{2.2\times10^{-4}} \right) 47 Gy =1.3Ty.
\label{tT}
\end{equation}

\bigskip

\noindent
From the result Eq.(\ref{tT}), we can find the number, defined 
in Eq.(\ref{NTCBE}), of causal patches at turnaround
\begin{equation}
N_T \simeq 2 \times 10^{122},
\label{NT}
\end{equation}
which is a very large number, as required.
From the result Eq.(\ref{tT}), we can also compute the number $E_T$
of necessary e-foldings between the present time and the turnaround
time
\begin{equation}
E_T = H^{-1} ( t_T - t_0) \simeq 94.
\label{ETCBE}
\end{equation}

\bigskip

\noindent
Based on the swampland conjectures, according to Eq.(\ref{ETstring}), the number of allowed
e-foldings before the SCs become violated is $E_T = O(1)$,
which is in tension with  Eq.(\ref{ETCBE}). 
To understand better this apparent disagreement, we can repeat our
calculations for two O(1) values of $E_T$:  $E_T = 2$ and $E_T= 4$. 
These values lead to smaller values for the
turnaround time
$t_T \sim 41$Gy and $t_T\sim69$Gy and very much smaller
values for the number of causal patches at
turnaround $N_T \sim 6,200 $ and $N_T \sim 2.3\times10^6$, respectively,
numbers which are far too small, see Eq.(\ref{NT}), for cyclic cosmology
to be possible.
We are employing general relativity at all times except
incrementally close to the turnaround and bounce, and do not expect
short times with unknown mathematics to change our conclusions.

\section{Discussion}

\noindent
To confirm directly that string theory is the correct theory of quantum gravity
is not yet possible because of the absence of relevant experimental
or observational data. This situation seems unlikely to change soon,
although even the best can underestimate future advances in technology.
Early last century Einstein thought experiments to measure classical gravitational
microlensing or gravitational waves were forever impracticable. Nevertheless,
in 2000\cite{Alcock} and 2016\cite{LIGO}, respectively, both have been
accurately measured.

\bigskip

\noindent
The swampland conjectures \cite{Vafa2005,OoguriVafa2006,OOSV,OPSV} about string theory
are extremely interesting because they are so extraordinarily strong in limiting the
allowed low-energy effective field theories. Our understanding is that they
were conjectured on the basis of studying many string theory models,
but are not proven to be necessary. While they
may play a role in confirming string theory, they also offer the
possibility that it might be refuted. We have discussed the issue of the future
of the universe and uncovered an apparent contradiction.

\bigskip

\noindent
To resolve this contradiction, we may entertain three explanations.
One possibility is the SCs be weakened relative to those
in the literature. Our Eq.(\ref{Es}) suggests a refined
version of SC1 where the range traversed by $\phi$ in field space is bounded by
$\Delta \sim O(100)$, instead of by $\Delta \sim O(1)$, in reduced Planck units.
A second possibility, difficult to believe, is that cyclic cosmology vastly overestimates
the number of e-foldings.
A third possibility, which is the least disfavoured, is that string theory is 
not the correct theory of quantum gravity.
Time will tell which of these possibilities is the truth.

\end{document}